\documentclass[linenumbers]{aastex631}
\usepackage[flushleft]{threeparttable}
\usepackage{booktabs}
\usepackage{footmisc}
\usepackage{lineno}

\shorttitle{Collimation of the Triple Source in Serpens}
\shortauthors{Rodr\'iguez-Kamenetzky, A. et al.}

\graphicspath{{./}{figures/}}

\begin{document}
\nolinenumbers

\title{Resolving the collimation zone of an intermediate-mass protostar}


\author[0000-0002-4731-4934]{Adriana R. Rodr\'iguez-Kamenetzky}
\affiliation{Instituto de Radioastronom\'ia y Astrof\'isica (IRyA-UNAM),\\
Morelia, Mexico}

\author[0000-0003-2862-5363]{Carlos Carrasco-Gonz\'alez}
\affiliation{Instituto de Radioastronom\'ia y Astrof\'isica (IRyA-UNAM),\\
Morelia, Mexico}


\author[0000-0003-2737-5681]{Luis Felipe Rodr\'iguez Jorge}
\affiliation{Instituto de Radioastronom\'ia y Astrof\'isica (IRyA-UNAM),\\
Morelia, Mexico}

\author[0000-0002-2110-1068]{Tom P. Ray}
\affiliation{Dublin Institute for Advanced Studies, Ireland}

\author[0000-0001-7960-4912]{Alberto Sanna}
\affiliation{INAF, Osservatorio Astronomico di Cagliari, via della Scienza 5, 09047, Selargius, Italy}
\affiliation{Max-Planck-Institut für Radioastronomie (MPIfR), Bonn, Germany}

\author[0000-0002-8517-8881]{Luca Moscadelli}
\affiliation{INAF, Osservatorio Astrofisico di Arcetri, Largo E. Fermi 5, 50125 Firenze, Italy}

\author{Melvin Hoare}
\affiliation{School of Physics and Astronomy, University of Leeds, Leeds, UK}

\author[0000-0003-1480-4643]{Roberto Galv\'an-Madrid}
\affiliation{Instituto de Radioastronom\'ia y Astrof\'isica (IRyA-UNAM),\\
Morelia, Mexico}

\author[0000-0001-8385-9838]{Hsien Shang}
\affiliation{Institute of Astronomy and Astrophysics, Academia Sinica, Taipei 10617, Taiwan}

\author[0000-0002-2260-7677]{Susana Lizano}
\affiliation{Instituto de Radioastronom\'ia y Astrof\'isica (IRyA-UNAM),\\
Morelia, Mexico}

\author[0000-0001-6496-0252]{Jochen Eisl\"offel}
\affiliation{Th\"uringer Landessternwarte, Sternwarte 5, D-07778 Tautenburg, Germany}

\author[0000-0003-4220-2404]{Jeremy Lim}
\affiliation{Department of Physics, The University of Hong Kong, Pokfulam Road, Hong Kong}

\author{Jos\'e M. Torrelles}
\affiliation{Institut de Ciències de l'Espai (ICE, CSIC), Can Magrans s/n, 08193, Cerdanyola del Vallès, Barcelona, Spain}
\affiliation{Institut d'Estudis Espacials de Catalunya (IEEC), Barcelona, Spain}

\author[0000-0002-3412-4306]{Paul Ho}
\affiliation{East Asian Observatory, 660 N. A'oh\={o}k\={u} Place, Hilo, Hawaii, 96720, USA}
\affiliation{Institute of Astronomy and Astrophysics, Academia Sinica,  Taipei 10617, Taiwan}

\author[0000-0002-3003-7977]{Anton Feeney-Johansson}
\affiliation{Dublin Institute for Advanced Studies, Astronomy \& Astrophysics Section, 31 Fitzwilliam Place, Dublin, D02 XF86, Ireland}
\affiliation{School of Physics, Trinity College Dublin, College Green, Dublin 2, Ireland}




\begin{abstract}
\nolinenumbers
We report new VLA and e-MERLIN high resolution and sensitivity images of the Triple Radio continuum Source in the Serpens star forming region. These observations allowed us to perform a deep multi-frequency, multi-epoch study by exploring the innermost regions ($\lesssim$100~au) of an intermediate-mass YSO for the first time, with a physical resolution of $\sim$15~au. The kinematic analysis of knots recently ejected by the protostar indicates that the jet is undergoing episodic variations in velocity. In addition, our multi-frequency images reveal striking characteristics, e.g., a highly collimated ionized stream that would be launched at a radial distance of $\sim$0.4~au from the protostar, and a narrow ($\sim$28~au wide) ionized cavity that would be excited by the interaction of a wide-angle component with the surrounding toroid of infalling material. In light of these results, we propose the scenario in which both a highly-collimated jet and a wide-angle wind coexist to be the most plausible to explain our observations, either launched by the X-wind or X- plus Disk-wind mechanism.
\end{abstract}

\keywords{stars: protostars --- stars: jets --- ISM: jets and outflows --- radio continuum: ISM}


\section{Introduction} \label{sec:intro}

Astrophysical jets are present in a wide variety of systems and play a crucial role in their evolution. However, a fundamental understanding of how they work has not been achieved, i.e., how they are launched and collimated. Moreover, it is still unknown whether or not there is a universal mechanism capable of explaining the origin of all types of jets. In this regard, protostellar jets, although being one of the least energetic manifestations of this phenomenon, constitute excellent touchstones to shed light on key questions that still await answers, mainly due to their proximity, which allows to probe close to their launching platforms.

Several models have been proposed to explain how protostellar jets are powered. The most commonly invoked group can be referred to as ``self-collimation'' models, which entail plasma confinement by a self-generated helical magnetic field in the protostar/disk system, similar to that proposed to explain relativistic jets \citep[e.g.,][]{Blandford1982}. Regarding self-collimation, two main scenarios can be distinguished depending on the distance from the protostar where ionized gas is launched and accelerated by magneto-centrifugal forces referred to as the X-wind and the Disk-wind frameworks. In the X-wind scenario, winds emerge from the gravitational X-point region, i.e., where the stellar magnetosphere intersects the disk, which co-rotates with the protostar \citep[e.g.,][]{Shu_1994,shu_2000, Shang_2007}. Here, the wind is launched on scales ranging from $\sim$0.03~au to $\sim$0.5~au from the source, and collimation into a jet is produced by density stratification and magnetic forces. In the Disk-wind scenario, winds are launched from the magnetized surface of the protostellar disk, which can occur close to the source but also up to a few au for an extended region along the disk \citep[e.g.,][]{Pudritz_Norman_1983,Pudritz_Norman_1986, konigl_2000,Pudritz_2007}. On the other hand, ``external collimation" mechanisms have also been proposed, where a poorly collimated or spherical wind can be collimated into a jet at distances of $\sim$100 au by a large-scale ordered magnetic field in the environment around the protostar \citep{Albertazzi_2014,ray_2021}. Moreover, different mechanisms could dominate the collimation of the material on different scales \citep{Frank_2014}, and even the possible dependence of the jet launching mechanism on the mass and/or evolutionary stage of the protostar might have to be taken into account \citep{Hoare_2015}.

Studying jet launching and collimation requires large observational efforts, mainly due to the small spatial scales involved and high sensitivity that is needed.
Recent studies of low-mass protostars at optical, millimeter and centimeter wavelengths revealed that collimation occurs below some tens of au \citep[e.g.][and references therein]{lee_2020, erkal_2021, Anglada_2018} along the jet axis, and the highest angular resolution observations of a low-mass protostar to-date (HL Tau) suggest that the jet is already collimated at distances of $\sim$1.5~au from the protostar \citep[Fig. 8 in][]{carrasco_2019}. Regarding the launching region, current evidence for low-mass protostars and T-Tauri stars (relatively evolved) is consistent with jets being launched from inner regions of the disk, supporting either X-wind or narrow Disc-wind models \citep[e.g.,][]{Lee_2008,Lee_2009,lee_2017,Liu_2012, erkal_2021}. There is also evidence of Disk-driven outflows in low-mass protostars \citep[e.g.,][]{Zhang_2018}, and recently, \citet{lee_2021} reported the first detection of interaction between a magnetic Disk-wind and an episodic jet in the low-mass jet system HH 212. Thus, observations of low-mass YSOs are currently consistent with a self-collimation scenario via either X- or Disk-wind models. On the other hand, high-mass protostars are located at much larger distances and therefore, the launching and collimation regions are more difficult to study. Observational evidence in high-mass star-forming regions indicate that wide-angle/spherical winds are commonly associated with massive protostars \citep[e.g.,][]{torrelles_1997,torrelles_2001,torrelles_2011,moscadelli_2007}, as well as poorly collimated large-scale molecular outflows \citep[e.g.,][]{arce_2007}; yet, highly-collimated jets powered by very young massive protostars have also been observed, e.g., Cep A HW 2 \citep[][]{curiel_2006,Carrasco_2021}, HH 80-81 \citep{rodriguez-kamenetzky_2017}, IRAS 16547-4247 \citep{lf_2005}, G035.02+0.35 \citep{sanna_2019}, IRAS 16562-3959 \citep{guzman_2010}, G016.59-0.05 \citep{moscadelli_2019}. Although magnetic fields are known to plays an important role in their collimation \citep[e.g.,][]{carrasco_2010, surcis_2014,sanna_2015,maud_2019}, it is still unknown whether the launching and collimation scales of highly collimated jets from massive protostars are as small as those observed in low-mass YSOs. Some works report evidence of Disk-driven rotating bipolar outflows, such as the case of the high-mass YSO candidate, Orion Source I \citep[e.g.,][]{hirota_2017}, while others suggest that the external medium can collimate the jet on larger scales, on the order of thousands of au from its driving source, regardless of the launching and initial collimation mechanisms \citep[e.g.,][]{ rodriguez-kamenetzky_2017}. Interestingly, collimation was found to occur on scales of tens of au from the W75N(B)-VLA 2 massive protostar, where the transition from an uncollimated wind to a collimated outflow due to its interaction with a toroidal environmental density stratification was observed \citep[e.g.,][]{carrasco_2015}. Additionally, in the last year, striking results have been found for the radio-jet powered by the massive protostar Cepheus A HW2 \citep{Carrasco_2021}. These authors mapped the outflowing material of this YSO below 100~au with a spatial resolution of $\sim$20~au, and discovered a morphology very different from that observed for low-mass protostars on similar scales. In order to explain their observations, two plausible scenarios were discussed: an extension of the classical Disk-wind to a massive protostar, and external collimation of a wide-angle wind. These evidences suggest that the dense environment in which massive stars are formed may also play an important role in jet collimation.

In this work we report new radio continuum images of the Triple Radio Source (TRS) in Serpens, which is a protostellar jet known to be driven by a Class 0 \citep[e.g.,][]{Hurt_1996,larsson_2000} intermediate-mass protostar \citep[$\sim$3 M$_\sun$,][]{Hull2017}, first noted in the far-infrared by \citet{Harvey_1984}, and commonly known as SMM1\footnote{SMM1 a.k.a Serpens FIRS1, SerpFIR1, Ser-emb 6, IRAS 18273+0113, S68 FIR, S68 FIRS1, and S68-1b.} according to its millimetric designation \citep{Casali_1993}. SMM1 is the brightest millimeter source in the Serpens Main star-forming region, located at a distance of 436$\pm$10 pc \citep{ortiz-leon_2017}, and was found to harbor a massive disk \citep[e.g.,][]{Hoger_1999,Enoch_2009}. 
The bolometric luminosity reported by \citet{kristensen_2012} for SMM1 is 30.4~L$_{\sun}$ (assuming a distance of 230 pc to the source), thus, scaling this value to the revised distance we obtain a bolometric luminosity of $\sim$ 100~L$_{\sun}$.
The radio continuum jet was first studied by \cite{LF_1989}, who named it after its morphology at radio frequencies, consisting of a central elongated thermal source and two external lobes (NW and SE) \citep[e.g.,][]{LF_1989,Curiel_1993,rodriguez-kamenetzky_2016}. Our new observations allow us to carry out a multi-epoch/multi-frequency study of the most deeply embedded regions ($\lesssim$ 100~au) of an intermediate-mass protostar for the first time, with a maximum spatial resolution of $\sim$15~au, leading to significant results regarding the star-formation process.

\section{Observations and Data Analysis} \label{sec:data}

The radio continuum emission of the TRS in Serpens was observed with the Karl G. Jansky Very Large Array (VLA) of the National Radio Astronomy Observatory (NRAO\footnote{The NRAO is a facility of the National Science Foundation operated under cooperative agreement by Associated Universities, Inc.}) and the enhanced Multi-Element Remotely Linked Interferometer Network (e-MERLIN\footnote{e-MERLIN is a National Facility operated by the University of Manchester at Jodrell Bank Observatory on behalf of STFC.}). In both cases the phase center was (J2000) 18$^h$29$^m$49.79$^s$ +01$^\circ$15$\arcmin$20.8$\arcsec$. A summary of the observations and their calibration process is provided in Table \ref{tab:obs} and following subsections.

\subsection{VLA data}\label{vla}

Multi-frequency observations were performed with the VLA using the A-configuration (VLA-A) at C (4-8 GHz), Ku (12-18 GHz), K (18-26.5 GHz), Ka (26.5-40 GHz), and Q (40-50 GHz) bands during December 2020 and January 2021 (see Table \ref{tab:obs}; Project code: 20B-122). We used the standard frequency setup for continuum observations, i.e., channels of 2 MHz width covering the entire frequency range at each band. Flux calibration was performed by observations of 3C286. For the higher frequency bands (K, Ku, Ka, and Q bands) we used J1743-0350 as our bandpass calibrator. For the C band observations, the flux calibrator 3C286 is bright enough to serve also as a bandpass calibrator. The complex gain (phase and amplitude) was corrected using frequent observations of the bright point-source quasar J1830+0619, with angular separation of $\sim$5~deg from the target source; we used the recommended calibrator-target-calibrator cycle times from the VLA observing manual (see table \ref{tab:obs}). Data calibration was performed by using the Common Astronomy Software Applications \citep[CASA,][]{CASA_NRAO} package (version 6.1.2.7) and the NRAO pipeline for VLA continuum observations. 

Images were made using the \emph{tclean} task of CASA. For each band, we made images with different visibility weightings (superuniform, natural and Briggs weighting with different values of the parameter \emph{robust}; see table \ref{tab:images} for the weighting used for each image shown). For all images we used multi-frequency synthesis with parameter \emph{nterms}=1.

\subsection{e-MERLIN data}\label{emerlin}

Observations at C band using the e-MERLIN interferometer were performed in two epochs: 2020 January and 2021 June (see Table \ref{tab:obs} for details on the observations). The central frequency was 5 GHz and we used a total bandwidth of 500 MHz with 1\,MHz channels. The flux and bandpass calibrators were 3C286 (same as in VLA observations) and J1407+2827, respectively. Phase calibration was performed by observing the bright quasar 1833+0115 every 3 minutes, which is separated from the target source by an angular distance of $\sim$0.8~deg. Calibration was performed by using CASA and the e-MERLIN pipeline. Images of the calibrated data were performed using \emph{tclean} in a similar way as in the case of  the VLA data.

In order to improve sensitivity, we also combined C-band e-MERLIN and VLA-A data. Since the observation dates differ from each other by several months (see Table \ref{tab:obs}), we applied a shift to the VLA-A uv-data in order to remove potential displacements of the whole region and possible errors due to the fact that we used different phase calibrators. With this purpose we first made separated C-band images from e-MERLIN and VLA-A data. We explored different weightings in order to obtain the best compromise between resolution and sensitivity, while minimizing the difference in beam size corresponding to each set of observations. This was achieved with Briggs weighting using parameter \emph{robust} = 0 and \emph{superuniform} weighting for the e-MERLIN and VLA-A data, respectively. Subsequently, both images were convolved to the same restoring beam, and re-gridded to the same dimension and pixel scale. Then, we computed the emission peak position through a Gaussian fit to each image, and calculated the offset between them. Taking into account this offset, 
we ran the {\tt\string fixvis} task of CASA to apply a shift to the VLA-A uv-data, in order to get both VLA-A and e-MERLIN data sets aligned. Finally, these data sets were concatenated in the same range of frequency (4.8-5.3 GHz) with the CASA task {\tt\string concat}, to produce a combined 500~MHz bandwidth image. The resulting image shows a morphology very similar to that observed with e-MERLIN alone but with less noise, as expected, and allowed us to recover some of the extended emission from the jet, as can be seen in the following sections.

\begin{deluxetable*}{lcccccccc}
\tablecaption{Observations}
\tablewidth{0pt}
\tablehead{
\colhead{} \vspace{-0.2cm} &\colhead{} & \colhead{} &  \colhead{} & \colhead{Effective Frequency} & \colhead{uv-range} & \multicolumn{3}{c}{Calibrators} \\ \cmidrule{7-9}
\colhead{Observation Date} \vspace{-0.2cm} & \colhead{Epoch} & \colhead{Interferometer} & \colhead{Band} & \colhead{Range [GHz]} &\colhead{[k$\lambda$]} & \colhead{Flux} & \colhead{Phase; Time cycle [min]} & \colhead{Bandpass}
}
\decimalcolnumbers
\startdata
2020 Jan 23, 25-27 & 2020.1 & e-MERLIN & C  & 4.82-5.33 & 60-3820 & 3C286 & 1833+0115; 3 & J1407+2827\\
2020 Dec 15 & 2021.0 & VLA-A & C  & 4.04-8.02 & 75-827 & 3C286 & J1830+0619; 8 & 3C286 \\
2020 Dec 26 & 2021.0 & VLA-A & K  & 17.98-26.02 & 30.5-3160 & 3C286 &  J1830+0619; 4 & J1743-0350 \\
2021 Jan 01 & 2021.0 & VLA-A &  Q  & 38.98-47.02 & 85-5740 & 3C286 &  J1830+0619; 2 & J1743-0350\\
2021 Jan 12 & 2021.0 & VLA-A & Ka  & 28.98-37.02 & 60-4500 & 3C286 &  J1830+0619; 3 & J1743-0350 \\
2021 Jan 15 & 2021.0 & VLA-A & Ku  & 12.04-18.02 & 20.5-2148 & 3C286 &  J1830+0619; 6 & J1743-0350 \\
2021 Jun 29,30 and Jul 01 & 2021.5 & e-MERLIN  & C & 4.82-5.33 & 60.2-3820 & 3C286 & 1833+0115; 3 & J1407+2827\\
\enddata
\end{deluxetable*}\label{tab:obs}

 \begin{deluxetable*}{lcccccc}
\tablecaption{Parameters of the images}
\tablewidth{0pt}
\tablehead{
\colhead{Band} \vspace{-0.2cm}  & \colhead{Interferometer} & \colhead{$\nu_{o}$ [GHz]} & \colhead{Weighting} & \colhead{Synthesized Beam; PA}  & \colhead{rms [$\mu$Jy/beam] } & \colhead{Used in}
}
\decimalcolnumbers
\startdata
 C & VLA-A & 6 & Robust 0.5 & 0$\farcs$443 $\times$ 0$\farcs$320; -63$^{\circ}$ &   10 &  Fig. \ref{fig:full_jet} (left) \\
  C$^{*}$ & VLA-B   & 6 & Natural & 0$\farcs$47 $ \times$ 0$\farcs$47 & 10 &  Table \ref{tab:pm} \\
  C & VLA-A+e-MERLIN   & 5 &  Robust 0 &  0$\farcs$118 $\times$ 0$\farcs$037; 25$^{\circ}$ & 5 & Fig. \ref{fig:full_jet} (right); Fig. \ref{fig:secuencia_TB} (a) \\
  C & e-MERLIN  & 5  &  Robust 0 &  0$\farcs$140 $\times$ 0$\farcs$050; 23$^{\circ}$ & 6 & Fig. \ref{fig:pm1} (upper panel) \\
  C & e-MERLIN  & 5  &  Natural &  0$\farcs$140 $\times$ 0$\farcs$050; 23$^{\circ}$ & 10 & Fig. \ref{fig:pm1} (bottom panel) \\
  Ku & VLA-A   & 15 &  Superuniform & 0$\farcs$107 $\times$ 0$\farcs$087; 13$^{\circ}$ & 6 & Fig. \ref{fig:secuencia_TB} (b) \\
  K & VLA-A   & 22 &  Robust 0 & 0$\farcs$082 $\times$ 0$\farcs$072; -27$^{\circ}$ & 6 & Fig. \ref{fig:secuencia_TB} (c) \\
  Ka & VLA-A   & 33 &  Superuniform & 0$\farcs$047 $\times$ 0$\farcs$044; -26$^{\circ}$ & 20 & Fig. \ref{fig:secuencia_TB} (d) \\
  Q & VLA-A  & 43  &  Superuniform & 0$\farcs$035 $\times$ 0$\farcs$034; -62$^{\circ}$ & 50 & Fig. \ref{fig:secuencia_TB} (e), (f)
\enddata
\tablecomments{$^{*}$ Reported in \cite{rodriguez-kamenetzky_2016}}
\label{tab:images}
\end{deluxetable*}

\section{Results} \label{sec:1}

\begin{figure}
    \centering
    \includegraphics[scale=0.6]{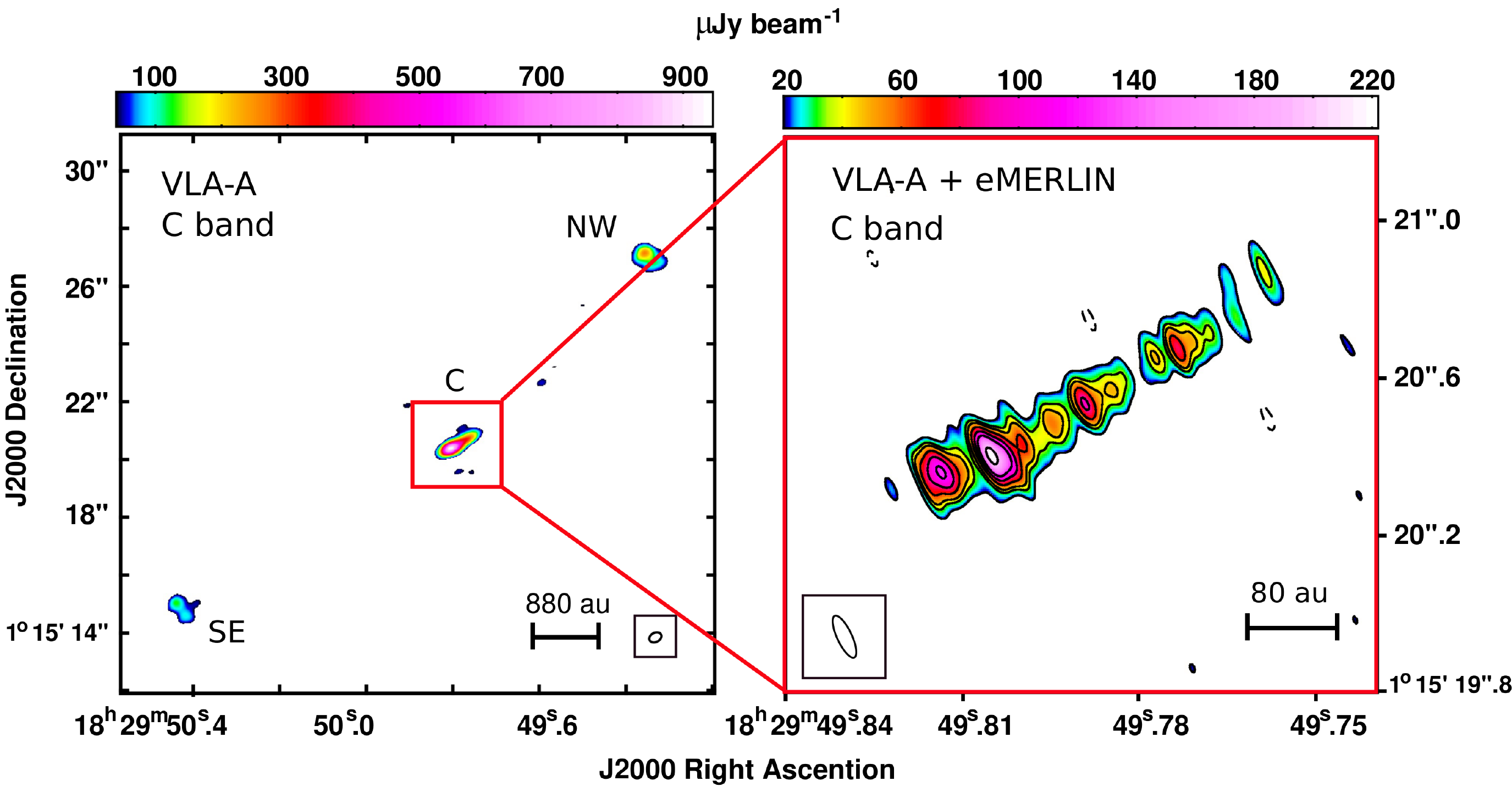}
    \caption{Radio continuum observations of the TRS in Serpens. Left panel: We show a VLA-A C band image above 4$\times$ the rms of 10~$\mu$Jy~beam$^{-1}$. The main sources detected are labeled (C, SE, and NW). Right panel: We show the deepest image of the Serpens jet core obtained by the combination of VLA-A and e-MERLIN data in the same frequency range (from 4.7 to 5.2 GHz). Contours are [-4, 4, 7, 9, 13, 18, 24, 40]$\times$ the rms of 5$\mu$Jy~beam$^{-1}$. Beam sizes and other parameters of the images are listed in Table \ref{tab:images}.
    }
    \label{fig:full_jet}
\end{figure}

In Figure \ref{fig:full_jet} (left panel) we present the continuum image of the jet obtained with the VLA-A at 6 GHz. As observed in previous works \citep[e.g.,][]{LF_1989,rodriguez-kamenetzky_2016}, we identify three main compact sources clearly manifesting the triple morphology associated with this YSO. The elongated structure at the center (source C) traces material recently ejected by the protostar. This emission can be described using the \citet{reynolds_86} models as free-free radiation ($-0.1<\alpha<+$0.6, with S$_{\nu}\propto\nu^{\alpha}$ the flux density at the frequency $\nu$) from a collimated ionized bipolar outflow, commonly referred to as a thermal radio jet \citep[e.g.,][]{rosero_2016,sanna_2018,Anglada_2018,kovak_2021}. The two outer knots, NW and SE, located at an angular distance of $\sim$7 arcsec from the protostar ($\simeq 3000$~au), are known to be diametrically moving away from the powering source of the jet displaying large proper motions, corresponding to tangential velocities between 200 and 300~km~s$^{-1}$ \citep[e.g.,][]{LF_1989,rodriguez-kamenetzky_2016}. These knots are likely tracing shocks of the jet with a dense ambient medium, and were found to be able to accelerate particles via the Fermi I mechanism under certain conditions \citep{rodriguez-kamenetzky_2016}. In such strong shocks the jet material is expected to slow down, implying pre-shock jet velocities higher than those observed. This means that the measured proper motions of the two outermost knots of the triple source do not reflect the true velocity of the jet before impact, but a lower limit. In Figure \ref{fig:full_jet} (right panel) we show a higher angular resolution image of the Serpens jet core (source C) obtained by the combination of VLA-A and e-MERLIN data in the same frequency range (see Section \ref{emerlin}, and Table \ref{tab:images}). This allows us to observe internal regions of the jet with an unprecedented angular resolution at these frequencies ($\sim$16 au, in the jet direction) and, at the same time, to recover extended emission. The resulting image reveals with great detail a chain of condensations even on scales smaller than 100~au from the protostar. Hereafter, we will refer to these condensations as the internal knots, to differentiate them from the outer sources NW and SE.

\subsection{Proper motions}\label{sec:pm}

\begin{figure}[ht!]
\includegraphics[scale=0.7]{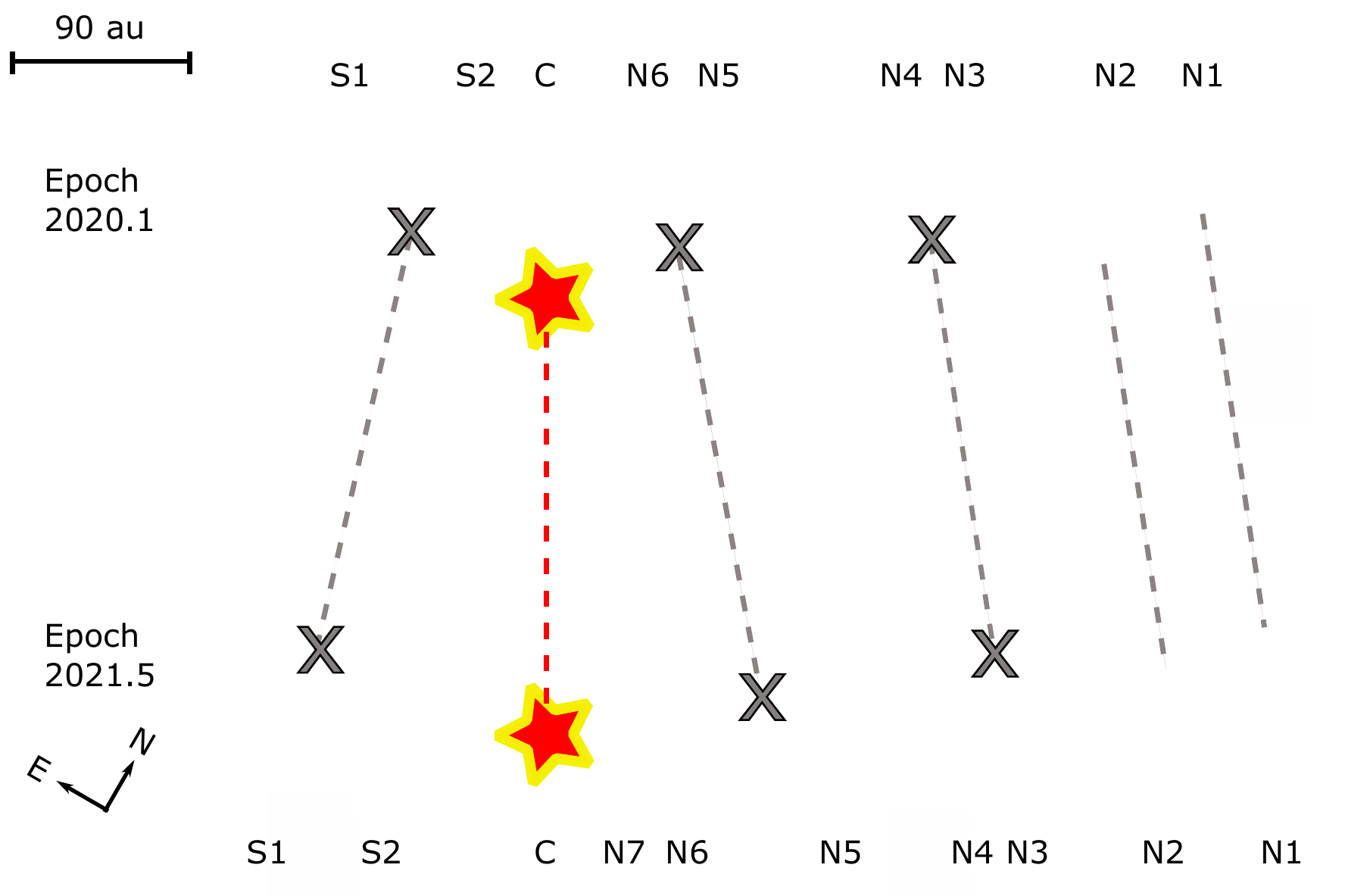}
\caption{Comparison of two e-MERLIN C-band images observed 1.4 years apart. Both images were convolved to the same beam size (0\farcs140 $\times$ 0\farcs050; PA 23$^{\circ}$) in order to measure the positions of the internal knots through a Gaussian fit to their intensity profiles. We label all of the knots that can be identified in both epochs above 4$\sigma$. Red stars indicate the position of the emission peak in the VLA-A Q-band image (Figure \ref{fig:secuencia_TB} panel e), considered to be the location of the powering source of the jet (indicated as C). Gray crosses indicate the averaged position of unresolved knots, i.e., S1\&S2, N5\&N6, and N3\&N4. The gray-dashed lines join the positions of the knots in the two epochs. EPOCH 2020.1: countours are [4, 5, 7, 8.5, 9.5, 10.4, 13, 15, 20, 25, 35, 40]$\times$ 6~$\mu$Jy beam$^{-1}$. EPOCH 2021.5: contours are [4, 5, 5.9, 6.7, 8, 10, 13, 15, 20, 30, 40]$\times$ 10~$\mu$Jy beam$^{-1}$.
 \label{fig:pm1}}
\end{figure}

In order to estimate proper motions of internal knots we compare C-band e-MERLIN images from the 2020.1 and 2021.5 epochs (see Figure \ref{fig:pm1}), which are convolved to the same beam size. A red star indicates the emission peak of a Gaussian fit to the Q-band VLA-A image, which has the maximum angular resolution achieved (see below, Figure \ref{fig:secuencia_TB} panel e; corresponding to the epoch 2021.0). As discussed in Section \ref{sec:zoom}, we consider this point to be the location of the powering source of the jet (indicated by the letter C). The uncertainty in its position is $\sim$5~mas ($\sim$2~au), given by the quadratic sum of the uncertainty in the Gaussian fit ($\sim$1~mas for RA and $\sim$2~mas for DEC) plus $\sim$4.5~mas corresponding to the absolute uncertainty of the Q-band VLA-A array observations ($\sim$10\% of the beam under typical conditions). Images in both panels of Figure \ref{fig:pm1} are of the same size and scale, and were aligned using the absolute position of C, as indicated by a red dashed-line in Figure \ref{fig:pm1}. In 2020.1 no clear emission associated with the position of the protostar is detected, in contrast to epoch 2021.5, where emission is detected tracing recently ejected material (within the previous year). We identify a chain of condensations which we label as S1, S2, N6, N5, N4, N3, N2, and N1. The distribution of knots on both sides of C are asymmetric. The series of internal knots N are relatively faint and extend over several hundreds of au. In contrast, there are just two bright S internal knots which only extend over 100 au (Figure \ref{fig:pm1}). This asymmetry suggests that the gaseous environment is denser to the SE than to the NW of the protostar. In addition, the direction of ejections to the SE and NW appear to be slightly misaligned, which could also be related to variations in ambient density. 

In order to study the kinematics of the jet core, we first compute the positions of the knots identified in both epochs (see Figure \ref{fig:pm1}) through a Gaussian fit to their intensity profiles. Positional errors are calculated taking into account Gaussian fitting errors and the astrometric accuracy of e-MERLIN observations (i.e., $\sim$1~mas at C-band). We measured the proper motions of knots N1 and N2 in a straightforward manner by computing their position in epoch 2021.5 relative to 2020.1. However, given that a number of knots are not resolved spatially in both epochs, and to avoid ambiguity in their identification, we calculated the displacement of their averaged positions between epoch 2020.1 and 2021.5, assuming these condensations preserve their identities. For example, we calculated the average position of knot S1 and knot S2 to provide a single position in each epoch, i.e., $\overline{\alpha}_I= (\alpha_{S1_I}+\alpha_{S2_I})/2$ and $\overline{\delta}_I= (\delta_{S1_I}+\delta_{S2_I})/2$ for the first epoch, and $\overline{\alpha}_{II}= (\alpha_{S1_{II}}+\alpha_{S2_{II}})/2$ and $\overline{\delta}_{II}= (\delta_{S1_{II}}+\delta_{S2_{II}})/2$ for the second one; thus, considering the averaged positions, we estimated the proper motion of the S1\&S2 condensation, and repeated this procedure for knots N5\&N6, and N3\&N4. The averaged positions are indicated in Figure \ref{fig:pm1} with gray crosses. We also estimate the tangential velocity, the PA of the movement, and the kinematic age, i.e., the time needed for them to move from the YSO (Q band emission peak) to their present position (2021.5), assuming constant velocity. Velocity errors are determined considering positional errors and the uncertainty in the distance to the region. Results are listed in Table \ref{tab:pm}, where we also include proper motions of the external knots, NW and SE (split into SE\_N and SE\_S), from 1993 to 2011 reported in \citet{rodriguez-kamenetzky_2016}. We notice that, within uncertainty margins, most of the condensations move with velocities of $\sim$100 km~s$^{-1}$. However, those more recently ejected seem to show a trend to increasing velocities going from $\sim$130~km~s$^{-1}$ to $\sim$160~km~s$^{-1}$. According to their kinematic ages (column 5 in Table \ref{tab:pm}), it seems that the jet had a velocity of $\sim$100~km~s$^{-1}$ for a period of $\sim$10 years, after which a slight transition to higher velocities began. In 2021.5 we also detect a new knot located between sources C and N6 (i.e., N7, see Figure \ref{fig:pm1}) that would have been ejected in the last year with an estimated tangential velocity $\gtrsim$120~km~s$^{-1}$.

\begin{deluxetable*}{lcccc}
\tablecaption{Proper motions}
\tablewidth{0pt}
\tablehead{
\colhead{Knot} & \colhead{$\mu$ [mas/yr]} & \colhead{V [km~s$^{-1}$]} & \colhead{PA [$^{o}$]} & \colhead{Kinematic Age [yr]}
}
\decimalcolnumbers
\startdata
S2\&S1 & 79 $\pm$ 6 & 160 $\pm$ 20 & 111 $\pm$ 6 & 4 $\pm$ 1\\
N5\&N6 & 61 $\pm$ 4  & 130 $\pm$ 10 & 290 $\pm$ 6 & 4 $\pm$ 1 \\  
N3\&N4 & 49 $\pm$ 7 & 100 $\pm$ 20 & 317 $\pm$ 8 & 10 $\pm$ 2\\  
N2 & 40 $\pm$ 10 & 90 $\pm$ 20 & 330 $\pm$ 10 & 16 $\pm$ 4 \\
N1 & 47 $\pm$ 5 & 100 $\pm$ 10 & 315 $\pm$ 7 & 18 $\pm$ 3\\
SE\_N$^{a}$ & 152 $\pm$ 1 & 299 $\pm$ 2 & 126 $\pm$ 1 & 63 $\pm$ 1 \\
NW$^{a}$ & 104 $\pm$ 1 & 205 $\pm$ 2 &  316 $\pm$ 1  & 80 $\pm$ 2 \\  
SE\_S$^{a}$ & 113 $\pm$ 1 & 222 $\pm$ 2 & 132 $\pm$ 1 & 85 $\pm$ 2
\enddata
\tablecomments{Parameters derived for the internal knots identified in Figure \ref{fig:pm1}, from 2020.1 to 2021.5. Parameters for knots KS, KN4, and KN3 are derived from averaged positions of the unresolved knots as indicated in column 1. $^{(a)}$ Parameters computed by \citet{rodriguez-kamenetzky_2016} for the NW and SE (SE\_N and SE\_S) external knots from 1993 to 2011; kinematic ages are estimated from the velocity of the knots and their respective distances to the central source, reported in the cited work. PAs are measured from North to East.}
\label{tab:pm}
\end{deluxetable*}


\subsection{Zooming into the innermost regions}\label{sec:zoom}

Here we perform a multi-frequency study which allows us to observe the deepest regions of this YSO, revealing that the radio continuum distribution near source C changes with frequency, from being aligned to the jet direction below about 30\,GHz, to emitting along a direction approximately perpendicular to the jet above 30 GHz (see Figure \ref{fig:secuencia_TB}). All images were made from VLA-A data, with the exception of panel (a), which corresponds to the combination of e-MERLIN and VLA-A data presented in Figure \ref{fig:full_jet} (see Table \ref{tab:images} for details). Upper panels show a sequence of increasing frequency images covering the same region of the sky: 5 GHz (panel a), 15 GHz (panel b), and 22 GHz (panel c). Lower panels in Figure \ref{fig:secuencia_TB} show a zoom at different frequencies of the region enclosed by a red rectangle in upper panels (33 GHz in panel d, and 43 GHz in panel e). The knotty structure of the radio jet is observed with different angular resolution from 5 to 22\,GHz, with an approximate overall length of 500 au (panels a, b, and c). We indicate the direction of the NW- and SE-jet counterparts with black-dashed arrows in panel (b) for ease of comparison with images at other frequencies (panels d and e). In panel (c), the emission at 22 GHz also reveals an extended component, probably tracing the walls of a cavity oriented in the NW direction and here we have drawn a parabolic dashed curve to help visualise it\footnote{This curve does not represent a fit to the emission profile.\label{curve}}. This is in line with a previous study presented by \citet{Hull_2016}, who reported the first direct detection of an ionized outflow cavity via free–free emission at centimeter wavelength in this YSO; \citet{liang_2020} applied a wind-blown cavity model to the shape. \citet{Hull_2016} explore different scenarios that could lead to the ionization of the outflow cavity, and found the most plausible to be: (1) ionization via UV photons escaping from the accreting protostellar source, and (2) shock ionization produced by a precessing molecular jet impacting the walls, or a combination of the two. However, this parabolic-like feature might also corresponds to the interface where a wide-angle wind slams into the ambient medium, heating and piling up gas at the cavity walls. On the other hand, moving to higher frequencies, the contribution of free-free emission from the jet is much lower and only its most compact structures are detected, as can be seen in panels (d) and (e), where we also indicate the jet direction from panel (b) and the parabolic curve drawn in panel (c), for reference. In panel (d) we see that the emission at 33 GHz is clearly dominated by an extended component elongated in a direction approximately perpendicular to the jet, with a $\sim$40 au semi-major axis. This kind of structure is reminiscent of a protostellar disk, however, part of the emission seems to match with the parabolic dashed curve corresponding to the NW-oriented cavity detected in panel (c). Moreover, the highest angular resolution image we obtained ($\sim$15~au at 43 GHz, panel e) reveals a narrower parabolic-like component curved towards the SE direction as indicated by the green dashed-line\footref{curve}, which is observed to emerge very near to the emission peak ($\sim$28~au wide at the base, measured perpendicularly to the jet direction at the 9$\sigma$ contour in panels e and f). We notice that this emission extends up to $\sim$30~au in the SE direction, while a collimated stream seems to start at a distance of $\sim$60~au from its driving source (emission peak in panel e, indicated in panel f with a magenta cross). As jets are visible as a result of shocks, we interpret the emission peak of this stream as the place where the first substantial ionizing shock occurs. Interestingly, the brightness temperature map (panel f) corresponding to the 7~mm image (panel e) over 8$\sigma$ shows values well above 200~K, which are very high compared to the brightness temperature that would be expected at this frequency for thermal emission from a dusty-disk up to $\sim$40 au from the protostar \citep[e.g.,][]{carrasco_2019}. This strongly suggests that the parabolic-like structure observed at 43\,GHz is most probably dominated by free-free radiation from ionized gas.

\section{Discussion}\label{sec:2}

The results obtained in the kinematic and multifrequency study presented in previous sections lead us to the following discussion:

\subsection{Jet kinematics}\label{sec:kin}

Regarding the kinematics of the outer knots, NW and SE, proper motion studies spanning 18 years indicate they are moving away from the YSO with velocities of $\sim$300~km~s$^{-1}$ \citep{rodriguez-kamenetzky_2016, LF_1989}. As discussed in \citet{rodriguez-kamenetzky_2016}, the ejected material is expected to slow down in strong shocks with the dense ambient medium (as seems to be the case for NW and SE), thus implying jet velocities higher than the measured values. In contrast, the inner chain of condensations we detect is likely tracing internal shocks in the jet, in which case velocities derived from proper motion measurements can be considered as more closely following the average tangential jet velocity \citep[e.g.,][]{marti_1995, LF_2000}. These internal knots are observed to move much slower than NW and SE, showing velocities of approximately 100 km~s$^{-1}$ with a slight trend to higher values for more recent ejections (Table \ref{tab:pm}). Periodic velocity variations are frequently invoked to explain the formation of knots in astrophysical jets, as emphasized by theoretical and numerical studies \citep[e.g.,][]{raga1990,romanova2009,romanova2016}. Thus, we propose that the internal chain of knots corresponds to a phase where the material is ejected with typical velocities of $\sim$100~km~s$^{-1}$, while the outer knots (NW and SE) were ejected during a more energetic stage (implying jet velocities higher than $\sim$300~km~s$^{-1}$) perhaps associated with an FU Ori-like outburst in the past \citep[e.g.][]{1996ARA&A..34..207H}. This periodicity could be caused by tidal interactions between the disk from which the jet originates and a binary star with an eccentric/non-coplanar orbit \citep[e.g.,][]{Masciadri_2002, Anglada_2007}. This is in agreement with the work of \citet{rodriguez-kamenetzky_2016}, who proposed that interactions of the driving source of the jet with a close ($\sim$3 au) binary companion could lead to episodic increases in ejecta velocities every 20-30 years, giving rise to strong shocks capable of accelerating particles to relativistic energies (e.g., NW and SE outer knots). According to this, we think that we could now be observing the transition from the slow to the fast ejection phase, and that the jet will become more energetic once more within approximately 10-20 years, exhibiting velocities higher than $\sim$300~km~s$^{-1}$. Since the jet suffers from precession, it is expected that not all of the new and faster ejections would intercept the slower material previously launched, and therefore, a larger number of epochs observed with comparable sensitivity would be valuable to perform a more detailed study of the core kinematics of this YSO.


\subsection{Jet collimation}\label{sec:coll}

Until recently, few YSOs have been resolved under 100 au at radio frequencies: e.g., the low-mass protostars DG Tau A \citep{Ainsworth_2013} and HL Tau \citep{carrasco_2019}, the low-mass multiple protostellar system L1551 IRS 5 \citep{Lim_2006, Lim_2016}, and the massive YSO Cep A HW2 \citep{Carrasco_2021}. In the case of DG Tau A, results are consistent with an outflow starting as a poorly focused wind which undergoes significant collimation farther along the jet ($\simeq$ 50 au), assuming it is launched very close to the protostar ($\simeq$1 au); see \citet{Ainsworth_2013}. On the contrary, in the case of L1551 IRS 5, studies have shown that jets collimate within 3 au of both binary components \citep{Lim_2006}, and the highest spatial resolution image obtained for HL Tau \citep{carrasco_2019} shows a very compact elongated source, suggesting that the jet is already collimated below 1.5 au from the protostar. This evidence is consistent with a self-collimation scenario where all the necessary conditions to launch and collimate the jet are created by the protostar/disk system itself. Finally, the massive YSO Cep A HW2 reveals another morphology yet, where two stationary elongated knots are detected either side of a compact spherical source, suggesting that collimation starts further away from the star (20-30 au). As previously mentioned (Section \ref{sec:1}), this suggests that the dense environment in which massive stars are formed may play an important role in jet collimation \citep{Carrasco_2021}.

\begin{figure}
    \centering
    \includegraphics[scale=0.8]{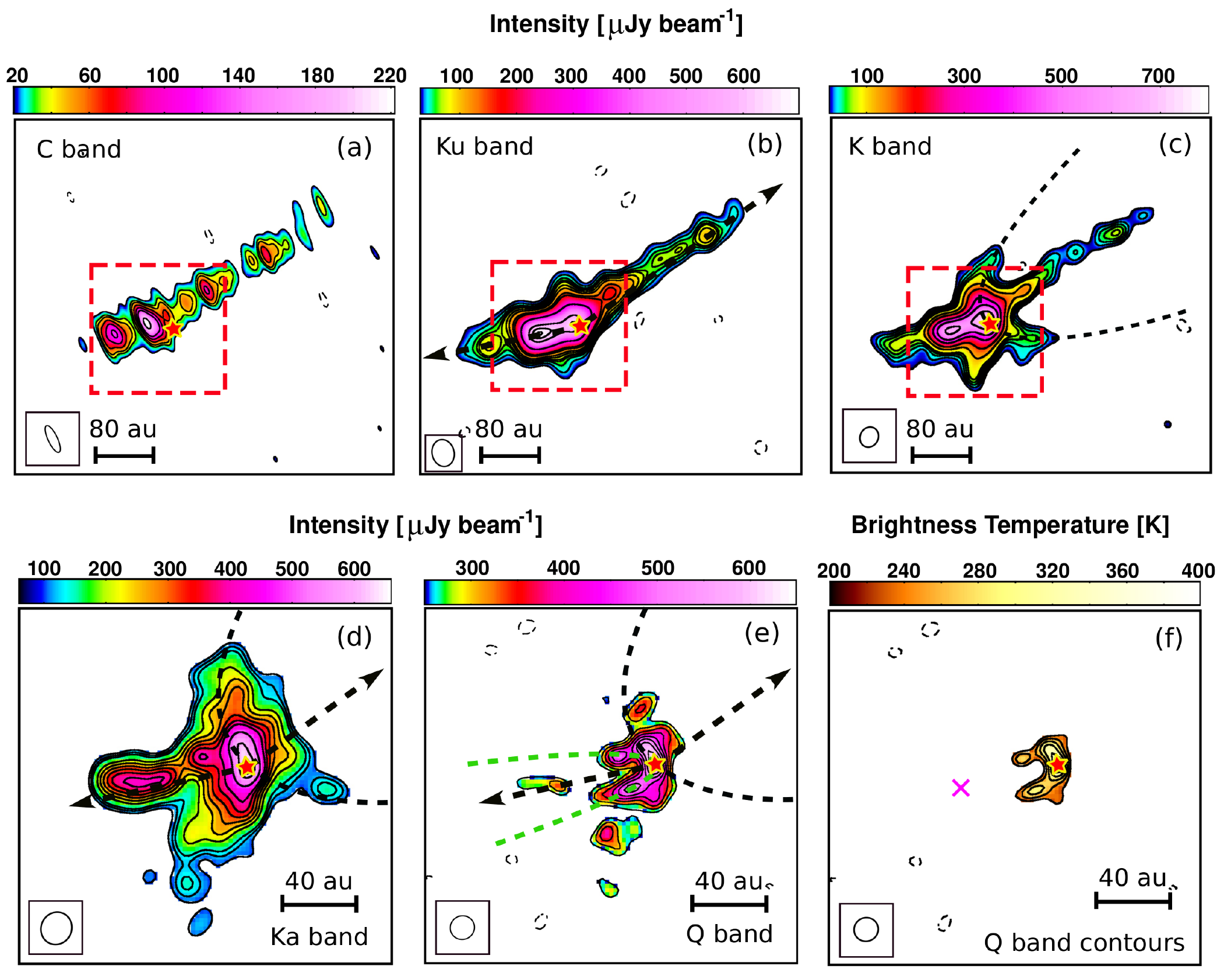}
    \caption{Images of the jet core at different frequencies. Upper panels show the same region of the sky, with the same physical scale. The red-dashed rectangles indicate the region surrounding the jet that is shown in the bottom panels (d, e, and f) with the same physical scale. A red star is shown in each panel to indicate the location of the emission peak in the highest angular resolution image obtained (panel e). (a) Briggs-weighted (Robust=0) C-band image (e-MERLIN and VLA-A data combined); countours are [-4, 4, 7, 9, 13, 18, 24, 40]$\times$ the rms of 5~$\mu$Jy beam$^{-1}$. (b) Superuniform-weighted image from Ku VLA-A data; contours are [-3, 5, 7, 9, 10.5, 11.5, 14, 17, 23, 27, 50, 90, 105, 113]$\times$ the rms of 6~$\mu$Jy beam$^{-1}$. NW- and SE-jet counterparts are indicated by black-dashed arrows.
    (c) Briggs-weighted (Robust=0) K-band VLA-A image; contours are [-4, 4, 6, 8, 10, 12, 20, 30, 50, 70, 100, 130]$\times$ the rms of 6~$\mu$Jy beam$^{-1}$. We draw a parabolic-dashed curve for descriptive purposes (see text). (d) Superuniform-weighted Ka band VLA-A; contours are [5, 6, 7, 10, 12, 14, 16, 18, 20, 25, 30, 32, 33]$\times$ the rms = 20~$\mu$Jy beam$^{-1}$. We also show the jet direction as identified in panel (b) and the black curve drawn in panel (c). (e) Q-band VLA-A superuniform-weighted image; contours are [-2, 5, 6, 7, 8, 9, 10, 11, 12]$\times$ the rms of 50~$\mu$Jy beam$^{-1}$. This is the highest physical resolution image achieved ($\sim$15 au). We draw a green parabolic-dashed curve for descriptive purposes (see text), in addition to the jet direction identified in panel (b) and the black curve drawn in panel (c). (f) Brightness temperature map corresponding to the 7~mm (Q band) emission (panel e) over 8 $\sigma$; contours are [-2, 8, 9, 10, 11, 12] $\times$ the rms of Q band emission. The magenta x-symbol in panel (e) indicates the position where the jet ionization seems to start.}
    \label{fig:secuencia_TB}
\end{figure}

Our work has shown striking characteristics of the innermost region ($\lesssim$100 au) of the intermediate-mass YSO of Serpens, revealing new details on the radio continuum nature, as shown in Figure \ref{fig:secuencia_TB}. We interpret these observations as evidence of two wind components with different collimation degrees, i.e., a highly collimated stream plus a wide-angle wind. The region of the disk from which they are launched cannot be clearly differentiated, however, since the terminal wind/jet velocity is a function of the Keplerian velocity ($v_{k,0}$) at the radius $r_0$ where the wind is launched \citep[e.g.,][]{Blandford1982,Shu_1994}, the high velocity measured for the internal knots in the jet (see Section \ref{sec:pm}) indicates that this particular component should be launched from an internal part of the disk. Presumably the outflow is mostly neutral, and what we detect is the emission from collisionally ionized gas. Thus, it is possible that the high-velocity/highly-focused stream (i.e., the jet) collimates very close to the driving source, but only manifests itself further out where shocks occur. Consequently, the compact elongated structure in the Ka- and Q-bands, which is located at $\sim60$~au from the protostar (see Fig. \ref{fig:secuencia_TB} d and e, respectively), could correspond to the first substantial internal shock detected at the observed epoch. Internal shocks travelling along the jet axis are frequently observed in astrophysical jets, and can be produced by a velocity variation at the base of the wind (accompanied or not with a mass-loss variation) \citep[e.g.,][]{canto_2000}. Such variations in velocity could result from, for example, tidal interactions of the disk with a binary companion or the presence of clumps during accretion. Likewise, for a steady X-wind, magnetic interactions with the ambient medium can produce magnetized internal shocks moving along the jet axis close to the wind speed \citep[][]{Shang_2020}; thus, the observed moving knots are also expected in the X-wind model. Another possibility is that the emission of the compact elongated structure is produced by a recollimation shock due to an external agent. As discussed in section \ref{sec:pm}, the velocity of the inner knots can be considered as more closely following the average tangential jet velocity \citep[e.g.,][]{marti_1995, LF_2000}, the values found being consistent with typical jet velocities predicted for X-winds ($\gtrsim$100~km~s$^{-1}$), which are characteristic of the potential well at the X-point \citep[][]{Shu_1994,Shu_1995}. Following \citet{Shu_1995}, the wind speed in the X-wind framework can be written as $v_w \sim 2 \sqrt{G M_* /r}$, thus, assuming a jet velocity of $\sim$160 km~s$^{-1}$ (as calculated in section \ref{sec:pm} for the S2\&S1 condensation), we estimate a launching point on the disk $r_X \sim 0.4$ au. New Q-band observations would indeed be valuable to follow up this shock in order to obtain a more accurate determination of the jet velocity, and also to study the possible appearance of new shocks closer to the protostar. On the other hand, the shell of ionized gas tracing a wide-angle morphology (see the parabolic feature indicated with a green-dashed line in Figure \ref{fig:secuencia_TB} (e)) could find a "natural" explanation in the interaction of a wide-angle wind component with the dense ambient medium in which the protostar is embedded (with an estimated envelope mass of $\sim$16~M$_{\sun}$ [\citealt{kristensen_2012}], and a high inner envelope density of $n(H_2)\gtrsim 5\times 10^{6}$~cm$^{-3}$ [\citealt{Goicoechea_2012}]). The X-wind, ejected from the X-point, is intrinsically a wide-angle magnetized wind with a jet-like component. The streamlines eventually collimate toward the rotation axis due to the toroidal component of the magnetic field. Nevertheless, the wind density distribution is cylindrically symmetric, thus, producing the “jet-like” appearance of the wind even at small distances from the protostar (see Figure 3 of \citet{Shu_1995} and Figure 4 of \citet{Shang_2002}). This wide-angle wind was predicted to be detected by the signatures of its collision with a flared accretion disk \citep[][]{li_1996} and/or the toroid that surrounds the YSO \citep{shang_2006,Shang_2020}. Given the pronounced curvature of the emission observed towards the SE direction in our Q-band image (see Figure \ref{fig:secuencia_TB}, panels e and f), we consider the interaction with a flared disk to be unlikely to produce the observed morphology. Thus, the X-wind model can produce both the jet and the wide-angle component, however, our observations do not allow to rule out the presence of winds ejected beyond the X-point, and a Disk-wind component must to be taken into account. Regarding Disk-winds, when the accretion rate is higher, the disk is heated further out from the protostar \citep[e.g.,][- expanding maser ring, and SED observations and radiative transfer modeling, respectively ]{Burns_2020,Stecklum_2021}; therefore, the frontier to which the disk is able to develop a wind should change depending on the accretion rate, moving back and forth. This implies that younger sources with higher accretion rates may be able to launch Disk-winds to farther out than older systems, which could also be detected through the interaction with the surrounding toroid. Thus, we speculate that the parabolic structure of ionized gas is tracing the interaction/collision of the wide-angled X-wind or a Disk-wind component with the infalling envelope in the form of a toroid. In light of this, the scenario in which both a highly collimated jet and a wide-angle wind coexist seems to be the most plausible way to account for the observations of the TRS in Serpens, either launched by the X-wind or X- plus Disk-wind mechanism. High angular resolution mm observations could be able to detect the dense environment shocked and heated by the wide-angle wind. In addition, both the slight misalignment between the NW and SE counterparts of the jet, as well as the observed asymmetry in the emission from the respective knots (Section \ref{sec:pm}), suggest that variations in the environmental density may also play a role in the focusing and orientation of the ejected material along with self-collimation mechanisms.

 Based in our study, it seems reasonable to consider a general picture for low/intermediate-mass protostars, in which deeply embedded YSOs undergoing intense accretion (Class 0) would be able to launch winds from different radii of the disk via both X- and Disk- wind mechanisms. In this picture, a highly collimated jet and structures similar to that observed in Q-band (Figure \ref{fig:secuencia_TB}e) can be expected, manifesting the presence of a wide-angle component through its interaction with the dense infalling envelope. In line with this, in more evolved YSOs (where accretion and environmental density have decreased), the hints of interaction of the wide-angle wind with the environment are expected to gradually fade away,
 remaining detectable only the faster and highly-focused winds launched from most internal regions of the disk, in agreement with observational evidences \citep[e.g.,][]{Liu_2012,erkal_2021}. On the other hand, as commented in Section \ref{sec:intro}, current evidence seems to favour a scenario where the dense environment in which massive stars form play an important role in the collimation of wide-angle winds at scales of tens of au \citep[e.g.,][]{carrasco_2015,Carrasco_2021}. Moreover, even when highly-collimated jets are observed up to thousands of au from massive protostars \citep[e.g., HH80-81][]{rodriguez-kamenetzky_2017}, it is not yet known whether they are as collimated as low/intermediate-mass protostellar jets at scales of few au. Thus, there is the possibility that different mechanisms could be at work to produce jets around most massive protostars stars. Future radio interferometers such as the Square Kilometer Array (SKA) and the Next Generation VLA (ngVLA) will allow to probe the launching regions with unprecedented angular resolution and sensitivity, and the synergy with numerical simulations on scales of few au would be valuable to shed light on this topic.

\section{Summary}\label{sec:summary}

 We have carried out a deep multi-frequency/multi-epoch study of the intermediate-mass protostar SMM1 in the Serpens star forming region, through high resolution and high sensitivity images obtained with the the VLA and e-MERLIN radio interferometers. These observations allowed us to resolve the radio emission of an intermediate-mass YSO within 100~au for the first time, unveiling important aspects of its nature.
 
 We resolved the core of the radio jet into a chain of condensations with a full detectable dimension of $\sim$500~au. Proper motion measurements show velocities of around 100~km~s$^{-1}$ for a period of $\sim$10 years, after which a slight trend to higher velocities ($\sim$200~km~s$^{-1}$) is observed. These internal knots are likely tracing the typical jet velocity, being much lower than values estimated for the external ($\sim$3000~au) shocks NW and SE observed in previous works. Our analysis supports the scenario in which the jet undergoes episodic increases in velocity every 20-30 years as proposed by \citet{rodriguez-kamenetzky_2016}. Moreover, this work demonstrates how e-MERLIN can be a valuable tool to study the innermost regions of young protostars, also showing the potentialities of future interferometers such as the SKA and the ngVLA.

 We also studied internal regions ($\lesssim$100~au) of the YSO at different frequencies, achieving an unprecedented spatial resolution ($\sim$15~au) at the highest frequency (43\,GHz), which allowed us to approach the collimation zone. The new images reveal the presence of a highly collimated ionized stream detected at a distance of $\sim$60~au from the protostar, and a narrow ($\sim$28~au wide) ionized cavity that would be excited by the interaction of either a wide-angled X-wind or a Disk-wind with the toroid of infalling material from the envelope that surrounds the protostar. According to our kinematic analysis, the highly-focused component (i.e., the jet) would be launched from internal regions of the disk at an estimated distance of $\sim$0.4~au from the protostar, considering the X-wind framework.
 \vspace{1cm}


\noindent{\it Acknowledgments}: This work was supported by UNAM DGAPA-PAPIIT
grants IN103921, IN108822 and IG101321, CONACyT Ciencia de Frontera grant number 86372. A.R.-K. thanks the UNAM DGAPA Postdoctoral Fellowship Program to support postdoctoral research. J.M.T. acknowledges partial support from the PID2020-117710GB- I00 grant funded by MCIN/AEI/10.13039/501100011033, and by the program Unidad de Excelencia María de Maeztu CEX2020-001058-M. T.P.R. acknowledges support from the European Research Council (ERC) under advanced grant number 743029. H.S. acknowledges grant support from Ministry of Science and Technology (MoST) in Taiwan through 109-2112-M-001-028- and 110-2112-M-001-019-. We would like to thank our referee for reviewing this article and providing constructive comments and suggestions.

\vspace{5mm}
\facilities{VLA, e-MERLIN}

\software{CASA}



\bibliography{sample631}{}
\bibliographystyle{aasjournal}



\newpage

\end{document}